\documentclass{elsarticle}
\usepackage{lineno,hyperref}
\modulolinenumbers[5]

\journal{International Journal of Heat and Mass Transfer}
\usepackage[utf8]{inputenc}
\usepackage{xcolor}
\usepackage{amsmath,amssymb}
\usepackage{bm}
\usepackage{graphicx}

\usepackage[normalem]{ulem}

\begin{document}

\newcommand{\EQ}{\begin{equation}}
\newcommand{\EN}{\end{equation}}
\newcommand{\EQA}{\begin{eqnarray}}
\newcommand{\ENA}{\end{eqnarray}}
\newcommand{\eq}[1]{(\ref{#1})}
\newcommand{\fig}[1]{fig.~(\ref{#1})}
\newcommand{\Eq}[1]{Eq.~(\ref{#1})}
\newcommand{\Eqs}[2]{Eqs.~(\ref{#1}) and~(\ref{#2})}
\newcommand{\eqs}[2]{(\ref{#1}) and~(\ref{#2})}
\newcommand{\Fig}[1]{figure~\ref{#1}}
\newcommand{\Tab}[1]{table~\ref{#1}}
\newcommand{\Figs}[2]{figures~\ref{#1} and \ref{#2}}
\newcommand{\Tabs}[2]{tables~\ref{#1} and \ref{#2}}
\newcommand{\mean}[1]{\overline #1}
\newcommand{\ddt}[1]{\frac{d#1}{dt}}
\newcommand{\bra}[1]{\langle #1 \rangle}
\def\Rey{\mbox{\rm Re}}
\def\Pe{\mbox{\rm Pe}}
\newcommand{\Nu}{\mathrm{Nu}}
\def\St{\mbox{\rm St}}
\def\Sh{\mbox{\rm Sh}}
\def\Sc{\mbox{\rm Sc}}
\def\Da{\mbox{\rm Da}}
\def\Kn{\mbox{\rm Kn}}
\def\Str{\mbox{\rm Str}}
\def\mp{m_{\rm p}}
\def\alphap{\alpha_{\rm p}}
\def\Vc{V_{\rm c}}
\def\Lf{L_{\rm f}}
\def\tauL{\tau_{\rm L}}
\def\tauc{\tau_{\rm c}}
\def\cs{c_{\rm s}}
\def\kf{k_{\rm f}}
\def\Cg{C_{\rm g}}
\def\Np{N_{\rm p}}
\def\np{n_{\rm p}}
\def\rhop{\rho_{\rm p}}
\def\urms{u_{\rm rms}}
\def\taup{\tau_{\rm p}}
\def\Xk{\bm{X}^{\rm k}}
\def\Vk{\bm{V}^{\rm k}}
\def\Deltax{\Delta x}
\def\Deltay{\Delta y}
\def\Deltaz{\Delta z}
\def\mp{m_{\rm p}}
\def\rp{r_{\rm p}}
\def\CD{C_{\rm D}}
\def\th{X}
\def\ths{X_{\rm s}}
\def\lambdat{\tilde{\lambda}}
\newcommand{\im}{\mathrm{i}}
\newcommand{\ex}[1]{\mathrm{e}^{#1}}
\newcommand{\du}{\mathrm{d}}
\newcommand{\Du}{\mathrm{D}}
\newcommand{\del}{\partial}
\newcommand{\Sec}[1]{\S\,\ref{#1}}
%
%
\newcommand{\ve}[1]{\boldsymbol{#1}}
\newcommand\vect[1]{{\mbox{\boldmath $#1$}}}
\newcommand{\gggg}{\mbox{\boldmath $g$} {}}
\newcommand{\ddd}{\mbox{\boldmath $d$} {}}
\newcommand{\xx}{\mbox{\boldmath $x$} {}}
\newcommand{\yy}{\mbox{\boldmath $y$} {}}
\newcommand{\zz}{\mbox{\boldmath $z$} {}}
\newcommand{\uu}{\mbox{\boldmath $u$} {}}
\newcommand{\rr}{\mbox{\boldmath $r$} {}}
\newcommand{\vv}{\mbox{\boldmath $v$} {}}
\newcommand
{\ww}{\mbox{\boldmath $w$} {}}
\newcommand{\mm}{\mbox{\boldmath $m$} {}}
\newcommand{\nn}{\mbox{\boldmath $n$} {}}
\newcommand{\PP}{\mbox{\boldmath $P$} {}}
\newcommand{\QQ}{\mbox{\boldmath $Q$} {}}
\newcommand{\UU}{\mbox{\boldmath $U$} {}}
\newcommand{\VV}{\mbox{\boldmath $V$} {}}
\newcommand{\bb}{\mbox{\boldmath $b$} {}}
\newcommand{\bp}{\mbox{\boldmath $p$} {}}
\newcommand{\qq}{\mbox{\boldmath $q$} {}}
\newcommand{\BB}{\mbox{\boldmath $B$} {}}
\newcommand{\HH}{\mbox{\boldmath $H$} {}}
\newcommand{\II}{\mbox{\boldmath $I$} {}}
\newcommand{\AAA}{\mbox{\boldmath $A$} {}}
\newcommand{\aaaa}{\mbox{\boldmath $a$} {}}
\newcommand{\jj}{\mbox{\boldmath $j$} {}}
\newcommand{\JJ}{\mbox{\boldmath $J$} {}}
\newcommand{\cc}{\mbox{\boldmath $c$} {}}
\newcommand{\eee}{\mbox{\boldmath $e$} {}}
\newcommand{\sss}{\mbox{\boldmath $s$} {}}
\newcommand{\ff}{\mbox{\boldmath $f$} {}}
\newcommand{\RR}{\mbox{\boldmath $R$} {}}
\newcommand{\EE}{\mbox{\boldmath $E$} {}}
\newcommand{\FF}{\mbox{\boldmath $F$} {}}
\newcommand{\TT}{\mbox{\boldmath $T$} {}}
\newcommand{\CC}{\mbox{\boldmath $C$} {}}
\newcommand{\KK}{\mbox{\boldmath $K$} {}}
\newcommand{\MM}{\mbox{\boldmath $M$} {}}
\newcommand{\GG}{\mbox{\boldmath $G$} {}}
\newcommand{\kk}{\mbox{\boldmath $k$} {}}
\newcommand{\hh}{\mbox{\boldmath $h$} {}}
\newcommand{\ee}{\mbox{\boldmath $e$} {}}
\newcommand{\SSS}{\mbox{\boldmath $S$} {}}
\newcommand{\grav}{\mbox{\boldmath $g$} {}}
\newcommand{\nab}{\mbox{\boldmath $\nabla$} {}}
\newcommand{\OO}{\mbox{\boldmath $\Omega$} {}}
\newcommand{\oo}{\mbox{\boldmath $\omega$} {}}
\newcommand{\ttau}{\mbox{\boldmath $\tau$} {}}
\newcommand{\ssigma}{\mbox{\boldmath $\sigma$} {}}
\newcommand{\PPi}{\mbox{\boldmath $\Pi$} {}}
\newcommand{\LLambda}{\mbox{\boldmath $\Lambda$} {}}
\newcommand{\llambda}{\mbox{\boldmath $\lambda$} {}}
\newcommand{\PPhi}{\mbox{\boldmath $\Phi$} {}}
\newcommand{\pomega}{\mbox{\boldmath $\varpi$} {}}
%
%
\newcommand{\KKKK}{\mbox{\boldmath ${\sf K}$} {}}
\newcommand{\JJJJ}{\mbox{\boldmath ${\sf J}$} {}}
\newcommand{\OOOO}{\mbox{\boldmath ${\sf O}$} {}}
\newcommand{\PPPP}{\mbox{\boldmath ${\sf P}$} {}}
\newcommand{\SSSS}{\mbox{\boldmath ${S}$} {}}
\newcommand{\TTTT}{\mbox{\boldmath ${\sf T}$} {}}
\newcommand{\AAAAA}{\mbox{\boldmath ${\sf A}$} {}}
\newcommand{\IIII}{\mbox{\boldmath ${\sf I}$} {}}
\newcommand{\LLLL}{\mbox{\boldmath ${\sf L}$} {}}
\newcommand{\MMMM}{\mbox{\boldmath ${\sf M}$} {}}
\newcommand{\BBB}{\mbox{\boldmath ${\cal B}$} {}}
\newcommand{\emf}{\mbox{\boldmath ${\cal E}$} {}}
\newcommand{\FFF}{\mbox{\boldmath ${\cal F}$} {}}
\newcommand{\GGG}{\mbox{\boldmath ${\cal G}$} {}}
\newcommand{\HHH}{\mbox{\boldmath ${\cal H}$} {}}
\newcommand{\QQQ}{\mbox{\boldmath ${\cal Q}$} {}}
\newcommand{\GGGG}{{\bf G} {}}

\newcommand{\yjour}[4]{, #2 {\bf #3}, #4 (#1).}
\newcommand{\pjour}[3]{, #2, in press (#1).}
\newcommand{\sjour}[3]{, #2, submitted (#1).}
\newcommand{\yprep}[2]{, #2, preprint (#1).}
\newcommand{\pproc}[3]{, (ed. #2), #3 (#1) (to appear).}
\newcommand{\yproc}[4]{, (ed. #3), pp. #2. #4 (#1).}
\newcommand{\ybook}[3]{, {\em #2}. #3 (#1).}

\newcommand{\yypr}[4]{, #3, pp. #2. #4 (#1).}

\def \d{{\rm d}}
\def \pd{{\partial}}
\def\nut{\nu}
\def\Rey{{\rm Re}}
\def\Le{{\rm Le}}
\newcommand{\frad}{\mbox{\boldmath ${\cal F}_{\rm rad}$}}
\newcommand{\fradscal}{{\cal F}_{\rm rad}}
\def\g{{$\gamma$}}
\def\nab{{\bm{\nabla}}}
\def\ff{{\bm f}}
\def\vq{{\bm q}}
\def\VV{{\bm V}}
\newcommand{\bez}{\begin{eqnarray*}}
\newcommand{\eez}{\end{eqnarray*}}
\newcommand{\be}{\begin{equation}}
\newcommand{\beq}{\begin{eqnarray}}
\newcommand{\eeq}{\end{eqnarray}}
\newcommand{\bc}{\begin{center}}
\newcommand{\ec}{\end{center}}
\newcommand{\cd}{carbon dioxide\xspace}
\newcommand{\hy}{hydrogen\xspace}

\begin{frontmatter}

\title{Thermophoresis and its effect on particle impaction on a cylinder for 
low and moderate Reynolds numbers}

\author[SINTEF,Nordita]{Nils Erland L. Haugen} 
\author[SINTEF]{Jonas Kr\"{u}ger}
\author[NTNU]{J{\o}rgen R. Aarnes}
\author[NTNU,Silesian]{Ewa Karchniwy}
\author[Silesian]{Adam Klimanek}

\address[SINTEF]{SINTEF Energy Research, N-7465 Trondheim, Norway}
\address[Nordita]{Nordita, KTH Royal Institute of Technology and Stockholm University, Roslagstullsbacken 23, SE-10691 Stockholm, Sweden}
\address[NTNU]{Department of Energy and Process Engineering, 
  Norwegian University of Science and Technology, 
  Kolbj{\o}rn Hejes vei 1B, NO-7491 Trondheim, Norway}
\address[Silesian]{Institute of Thermal Technology, Silesian University of Technology, Konarskiego 22, 44-100 Gliwice, Poland}

\date{}

\begin{abstract}
The effect of thermophoresis on the impaction of particles on a
cylinder is investigated for different particle sizes, particle
conductivities, temperature gradients and for
Reynolds numbers between 100 and 1600. Simulations are performed using
the Pencil Code, a high-order finite difference code. An overset-grid
method is used to precisely simulate the flow around the cylinder. The
ratio of particles impacting the cylinder and the number of particles
inserted upstream of the cylinder is used to calculate an impaction
efficiency.
It is found that both the particle conductivity and the temperature
gradient have a close to linear influence on the particle impaction
efficiency for small particles
Higher Reynolds numbers result in higher
impaction efficiency for middle-sized particles, while the impaction
efficiency is smaller for smaller particles.  In general, it is found
that thermophoresis only has an effect on the small particles, while
for larger particles the impaction efficiency is controlled by inertial
impaction.

Finally, an algebraic model, developed based on fundamental principles, which
describes the effect of thermophoresis is presented.
The model is found to accurately predict the DNS results. As such, this
model can be used to understand the mechanisms behind particle deposition
due to the thermophoretic force, and, more importantly, to identify
means by which the deposition rate can be reduced.
\end{abstract}


\end{frontmatter}


\section{Introduction}
Particle impaction on surfaces can be found in a multitude of
industrial systems, such as filters and heat exchanger surfaces. The
impaction and deposition of material on these surfaces can significantly
alter their performance, necessitating decreased maintenance intervals
or an increased rate of replacement of components. In order to improve
the design of surfaces exposed to particle laden flows, a thorough
understanding of the underlying effects is needed.
In this work, we will focus on how particles are transported to the
solid surface. For a particle to deposit on the surface, it must first
be transported to the surface before it has to stick to it. The latter mechanism
is outside the scope of this study, and here all particles impacting
on the surface will count towards particle deposition.

The transport of material to the surface is governed by the
impaction efficiency, which is the ratio of the number of particles that
actually come in contact with the cylinder to the number of particles
that would come in contact with the cylinder if they were to move in a
straight line, in the flow direction, from their point of origin.  For an overview of the
different approaches the reader is referred to the work of
\cite{Kleinhans2017}.

Due to its simplicity, a cylinder
placed in a particle laden flow is a widespread test case used to
study the impaction of particles on solid surfaces or heat exchanger
tubes. A sketch of such a case is shown in \Fig{fig:sketch},
where the particles (shown in green) are inserted from a plane (red),
which has the size of the projected area of the cylinder.
The initial velocity of the particles is equal to the flow velocity at the
insertion plane.
If the 
particles followed the flow from left to right, without any change in velocity,
all
particles would hit the cylinder, leading to an impaction efficiency
($\eta$) of unity. In reality, this does not happen as the fluid is
flowing around the cylinder, and particles are dragged along with
it. A particle's ability to follow the fluid is expressed as the
particle's Stokes number $\St$, which is the ratio of the particle
response time and the fluid time scale (details in
\Sec{sec:parteq}). In general, a particle with a Stokes number above
unity does not follow the flow very well, while particles with Stokes
numbers below unity tend to follow the flow they are embedded in.
\begin{figure}
 \centering
 \includegraphics[width=0.7\textwidth]{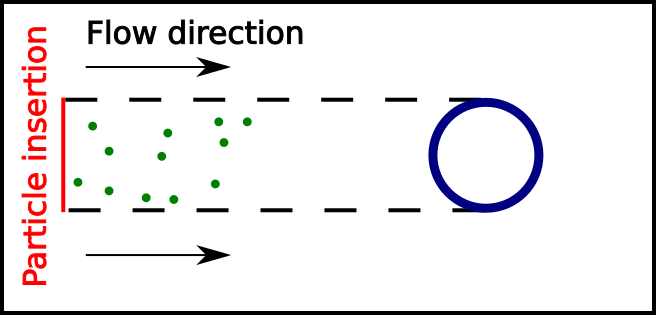}
 \caption{Sketch of particle deposition analysis case}
 \label{fig:sketch}
\end{figure}
\cite{Israel1982} developed a correlation based on potential flow
theory, which allows to calculate the impaction efficiency of
particles on an isothermal cylinder. This correlation is a well
established tool for predicting the isothermal impaction efficiency of
\emph{large} particles in a laminar flow, but does not yield correct
results for \emph{small} particles \citep{Haugen2010}.

The actual mass accumulation rate on a cylinder is determined by the
capture efficiency, which is the product of the impaction efficiency
and the sticking efficiency. The sticking efficiency is the
fraction of the impacting particles that actually stick to the surface
instead of rebounding. If either the particles or the cylinder surface
is at least partially melted, making them sticky, the sticking
efficiency is close to unity. On the other hand, for cold and clean
surfaces, particles will most likely bounce off the surface. The
sticking efficiency is then essentially zero. 

An experimental study, performed by
\cite{Kasper2009}, looked at the effect of mass accumulation on the
capture efficiency  and
proposes an empirical power law for it. Moreover, they present a new fit function
for the capture efficiency, which is bounded between 0 and 1.
The particle Stokes numbers in said study were between 0.3 and
3.
\cite{Haugen2010} investigated the impaction efficiency using Direct Numerical
Simulation with an immersed boundary method  and
found a steep drop in impaction efficiency below a certain Stokes number,
as particles become smaller and follow the flow better.  Extending
this work, Aarnes {\it et al.} studied the same case using overset grids,
obtaining results that are deemed more accurate with significantly
less computational efforts \citep{Aarnes2018,Aarnes2019}.

The effect of thermophoresis on the capture efficiency is studied by
several groups, both experimentally and numerically.
\cite{Beckmann2016} measured deposition of fly ash of a pulverized
coal jet flame and simulated the deposition rate using CFD and found
that thermophoresis increases the capture efficiency for smaller
particles, and that the relative increase is higher the smaller the
particles in question are.  Experimental data from a pilot-scale
furnace is compared to numerical results by \cite{Yang2017}, where the
influence of deposition growth on the particle impaction and sticking
efficiency is studied. They report that the higher surface temperature
due to deposit growth, results in a reduced effect of thermophoresis
and an increased sticking efficiency.  At later times, the rate of
shedding of material from the surface and deposition of material to
the surface from the flow balance out, so no net change of mass
sticking to the surface is observed \citep{Zhou2007}.

In the work of \cite{Kleinhans2019} the effect of the thermophoretic
force is studied both experimentally and numerically. In the
experimental part, the deposition of material on cooled and un-cooled
probes that are inserted into the particle laden flow above the burner
section of a combined heat and power (CHP) plant is
studied. Large-Eddy simulations are used to study the influence of the
sticking model and thermophoresis on deposition rate predictions. They
present a model that can take into account different sticking
mechanisms by which large and small particles of different composition
deposit. It is reported that thermophoresis accounts for three
quarters of the observed deposition rate.
\cite{GARCIAPEREZ2016408} used unsteady RANS simulations to study the effect
of thermophoresis on particle deposition and found that the thermophoretic
force was the dominating deposition mechamism for small particles.
To the knowledge of the current authors, Direct Numerical
Simulations (DNS) have not previously been used to perform a parameter
study of the effect of the thermophoretic force on the particle deposition
rate on a cylinder in a cross flow.
This further motivates the
authors of the current study to investigate the influence of different
flow conditions, such as flow Reynolds number, temperature gradient
and particle attributes on the effect of thermophoresis. This is the
aim of the present study.

\section{Theory}
The governing fluid equations are the ones for continuity, momentum
and energy. Pressure is taken into account through the ideal gas law
and the Mach number is $\sim 0.1$, which is so low that the flow is
considered as essentially incompressible.
\subsection*{Fluid equations}
The continuity equation is given by
\begin{align}
 \frac{\partial \rho}{\partial t} + \uu \cdot \nabla \rho &= - \rho \nabla 
\cdot \uu,
\end{align}
with $\rho$, $t$ and $\uu$ being density, time and velocity, respectively. The 
equation governing the conservation of momentum is
\begin{align}
 \rho \frac{\partial \uu}{\partial t} + \rho \uu \cdot \nabla \uu &= - \nabla p 
+ \nabla \cdot (2 \rho \nu S),
\end{align}
where $p$ is the pressure, $\nu$ the kinematic viscosity, and 
\begin{align}
 S &= \frac{1}{2} \left( \nabla \uu + (\nabla \uu)^{T} \right) - \frac{1}{3} I 
\nabla \cdot \uu
\end{align}
is the rate of strain tensor where $I$ is the identity matrix. 
The energy equation is
solved in the form of temperature:
\begin{align}
 \frac{\partial T}{\partial t} + \uu \cdot \nabla T &=  
\frac{k_f}{\rho c_{v}} \nabla^{2} T + 
\frac{2\nu S^{2}}{c_{v}}-\left(\gamma-1\right)T\nabla \cdot \uu,
\end{align}
where $\gamma=c_{p}/c_{v}=5/3$ and $c_{v}$ and $c_{p}$ are the heat capacities 
at 
constant volume and 
pressure, respectively, while $k_f$ is the thermal conductivity.
The ideal gas law is used to tie pressure and density together:
\begin{align}
 p &= \rho r_u T,
\end{align}
where $r_u=c_p-c_v$ is the specific gas constant. To simplify the
investigation, the kinematic viscosity, $\nu$, is assumed to be
constant since the temperature variations in the fluid are relatively
small.  Since the Pencil-Code is an explicit compressible code, the
time step is limited by the speed of sound through the CFL number. The
speed of sound is given by $c_s=\sqrt{\gamma r_u
  T}=\sqrt{c_p(\gamma-1)T}$. One can therefore use $c_p$ as a free
parameter in order to artificially lower the speed of sound to obtain
larger time steps. This is a valid approach as long as the Mach number
is kept lower than 0.1 and the viscous heating of the fluid is
negligible. To maintain a constant thermal diffusivity
($D_{\rm  thermal}=k_f/(\rho c_p)$) of the gas phase, and hence a constant
Prandtl number ($\Pr=\nu/D_{\rm thermal}$), the conductivity ($k_f$) is
changed proportionally to the specific heat capacity of the fluid
($c_p$).

\subsection{Particle equations}
\label{sec:parteq}
The particles considered here are spherical and have low Biot numbers,
which make them spatially isothermal. Numerically they are treated as
point particles that are influenced by the fluid, but are too
dilute to have any significant back-reaction on the fluid. In other words,
they are acted on by the flow but have no effect on it. This
assumption is applicable for dilute flows, which is the focus of the
current work.  The particle size is described by its Stokes number
\begin{align}
 \St = \frac{\tau_{St}}{\tau_{f}},
 \label{eq:St}
\end{align}
where $\tau_{St}= \frac{S_{\rho} d^{2}_{p}}{18\nu}$ is the particle Stokes 
time and $\tau_{f}=\frac{D}{u}$ is the flow time scale. Here, 
$S_{\rho}=\frac{\rho_p}{\rho}$ is the density ratio between particle and 
fluid, $d_{p}$ is the particle diameter and $D$ is the diameter of the cylinder.
Two forces are acting on the particle: the drag force and the 
thermophoretic force, while gravity is neglected for the small
particles studied here. 
The drag force is given by:
\begin{align}
\label{F_drag}
 \FF_{D} &= \frac{m_{p}}{\tau_{p}}\left(\uu-\vv_{p}\right),
\end{align}
where $\tau_{p}$, $m_p$ and $\vv_{p}$ are the particle's
response time, mass and velocity, respectively. Using the Stokes time with 
the Schiller-Naumann correction term \citep{Schiller1935} to account for low to 
moderate particle Reynolds numbers, the particle response time becomes:
\begin{align}
\label{eq:taup}
 \tau_{p} &= \frac{\tau_{St}}{f},
\end{align}
where
\EQ
\label{eq:corr_term}
f=1 + 0.15\Rey_{p}^{0.687}
\EN
and
$\Rey_{p} = d_{p}|\vv_{p}-\uu|/\nu$ is the particle Reynolds number.

The thermophoretic force pushes particles from regions of high
temperature to regions of low temperature. As such, it is similar to
the Soret effect for gases.  It was first observed in 1870 by
\cite{Tyndall1870}, and it has later become widely studied both
experimentally and theoretically. A theoretical analysis of the
thermophoretic force can be found in the works of
\cite{Zheng2002}. \cite{Young2011} gives an overview over the
different regimes of thermophoresis, which are determined by the
particles Knudsen number $\Kn=\lambda/d_{p}$, where $\lambda$ is the
mean free path of the gas. For the present study, all
particles are in the continuum regime, which is defined by $\Kn \ll 1$.
The thermophoretic force is then calculated by
\begin{align}
\label{F_th}
 \FF_{th} &= \Phi \frac{\mu^{2} r_{p} \nabla T}{\rho T},
\end{align}
where $r_p=d_p/2$ is particle radius, $\mu=\rho\nu$ is dynamic viscosity and
$\Phi$ is the thermophoretic force term. The expression
for $\Phi$ is taken from \cite{Epstein1929}:
\begin{align}
 \Phi &= \frac{-12 \pi K_{tc}}{2+\Lambda},
\end{align}
where the conductivity ratio between the particle and the gas
is given by $\Lambda=k_{p}/k_f$, while
the temperature creep coefficient, $K_{tc}$, used in this work has a
value of 1.1, which is in the middle of the range reported by
\cite{Sharipov2004}. This rather simple model for $\Phi$
simplifies the analysis, while still
providing agreeable results when compared with the widely used approach
proposed by \cite{Talbot1980}. From 
\Fig{fig:simple_vs_talbot} we see that the largest relative difference
between the simplified $\Phi$ and the one obtained when using the
approach of Talbot is less than a factor of two.
\begin{figure}
 \centering
\includegraphics[width=0.5\textwidth]{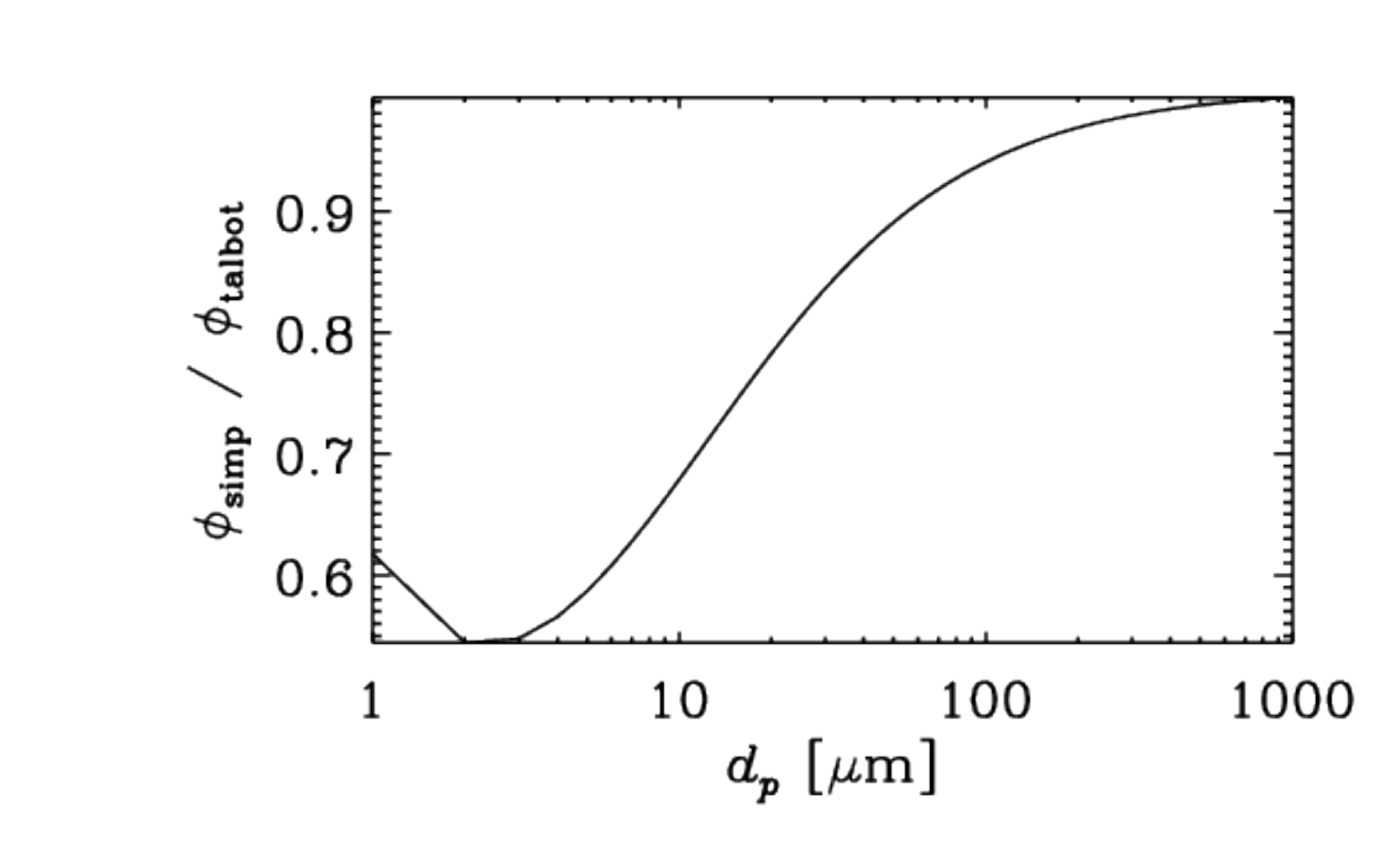}
 \caption{Comparison of different thermophoretic force terms.}
 \label{fig:simple_vs_talbot}
\end{figure}
\subsection{Theory}
Due to their short response times, very small particles will follow
the fluid almost perfectly, essentially behaving like tracer
particles. For isothermal situations, \cite{Haugen2010} showed that a
small fraction of these particles will nevertheless impact on the
cylinder surface due to their small but finite radii.

For the non-isothermal case, where the temperature of the cylinder is
lower than that of the surrounding gas, the thermophoretic force
will induce a relative velocity between the particles and the fluid
that transport the particles in the direction towards the cylinder.
The effect of this is that a larger fraction of the particles
impact on the cylinder surface. In the following we will try to quantify this
effect.

For laminar flows, a fluid
streamline that starts far upstream of the cylinder with a
displacement $\Delta x$ from the central line (the line that is
parallel to the mean flow and go through the center of the cylinder)
will move in the boundary layer of the cylinder with a radial
displacement from the cylinder surface of $\Delta r_f\ne \Delta x$
(see \Fig{fig:streamline}).
\begin{figure}
 \centering
 \includegraphics[width=0.7\textwidth]{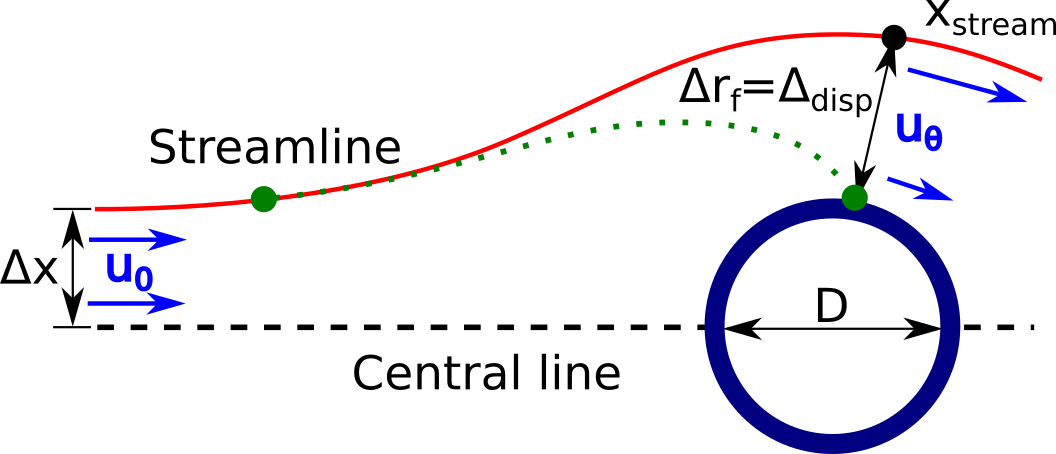}
 \caption{Sketch of a particle track of a particle with a small Stokes number 
   under the influence of thermophoresis.}
 \label{fig:streamline}
\end{figure}
Far upstream of the cylinder, the mass flow rate between a streamline and
the center line of the cylinder is given by
\EQ
\label{mdotu}
\dot{m}_u=Hu_0\rho_0\Delta x,
\EN
where $\Delta x$ is the distance between the streamline and the center line
and $H$ is the height of the cylinder. Within the boundary layer of the
cylinder, however, the mass flow rate between the streamline and the cylinder 
surface
is given by
\EQ
\dot{m}_b
\label{mdotb}
=\int_0^{\Delta r_f}\rho H u_\theta dr_{\rm cyl}
=\frac{\mean{\rho} H u_0\Rey^{1/2}}{2BD}(\Delta r_f)^2,
\EN
where $\mean{\rho}$ is the average fluid density within the boundary layer, and,
by following \cite{Haugen2010}, we have used 
	the fact that the tangential fluid 
velocity within the boundary layer is given by
\EQ
\label{utheta}
u_\theta(r_{\rm cyl})=\frac{u_0\Rey^{1/2}}{BD}r_{\rm cyl},
\EN
where $u_0$ is the far field fluid velocity,
$B$ is a constant of the order of unity and $r_{\rm cyl}$ is the normal distance from the cylinder surface.
Since the fluid is not turbulent (i.e., streamlines do not cross each other),
we know from mass conservation that
the mass flux between the streamline and the central line upstream of
the cylinder ($\dot{m}_u$) is equal to the mass flux between the
streamline and the cylinder surface ($\dot{m}_b$).  

The impaction efficiency for a stationary non-turbulent flow is given by
\EQ
\label{eta_small}
\eta=2 \Delta x_{\rm max}/D
\EN
where $\Delta x_{\rm max}$ is the maximum $\Delta x$ of the particles that
impact on the surface. Since streamlines of laminar flows do not cross each other, it is
clear that for a given type of particles, all particles inside $\Delta x_{\rm 
max}$
will impact on the cylinder surface while none of the particles outside of 
$\Delta x_{\rm max}$
will impact. For Reynolds numbers above $\sim 48$, the flow will become unsteady
and von Kármán eddies will occur in the wake of the cylinder.
The effect of this unsteadiness
on the front side impaction will be minor. Back side impaction may,
however,
be strongly affected by the von Kármán eddies. Since the focus of the current work
is to study front side impaction, the definition of the impaction efficiency as given
by \Eq{eta_small} will also be used for unsteady flows.

By calculating the distance a particle
will move in the radial direction due to the thermophoretic force
during the time it is in the front side
boundary layer of the cylinder, $\Delta_{\rm disp}$, and setting this
equal to $\Delta r_f$, the above equations can be solved to find the
impaction efficiency for small particles.
In the following, the focus will therefore be on finding $\Delta_{\rm disp}$.
Here, small particles are defined as particles that are so small that the
main cause of impaction is the thermophoretic force. This is typically the
case for $\St \lesssim 0.1$.

By combining \Eq{F_drag} and \Eq{F_th}, while setting the radial
component of the gas phase velocity 
to zero, we find the thermophoretic velocity of the particles (in the radial
direction) as
\EQ
\label{v_th}
v_{\rm th}=\frac{\Phi \nu}{6\pi f}\frac{\nabla T}{T}.
\EN
The correction term to the Stokes time, as
given by $f$ (see \Eq{eq:corr_term}), is always close to unity if the thermophoretic
force is the main driver of the particle velocity relative to the surrounding fluid.
We therefore set $f=1$ for the remainder of this analysis.
From Prandtl's concept of thin boundary layers, we know that the thickness
of the velocity 
boundary layer 
 can be approximated by
\EQ
\delta_{\rm vel}=\frac{D}{\Rey^{1/2}B}.
\EN
The thermal boundary layer thickness
is then given by \citep{Schlichting1979}
\EQ
\delta_{\rm thermal}=\delta_{\rm vel}{\Pr}^{-1/3}.
\EN
Hence, the average thermal gradient in the boundary layer 
becomes
\EQ
\label{nabla_T}
\nabla T\approx \frac{\Delta T}{\delta_{\rm thermal}}
=\frac{\Delta T}{D}B\Rey^{1/2}{\Pr}^{1/3}.
\EN
The effect of the thermophoretic force on the position of the particle
can now be considered as the radial
displacement of the particle ($\Delta_{\rm disp}$) from the position 
($x_{\rm stream}$) it would have 
without the influence of the thermophoretic force (see
\Fig{fig:streamline}). 
This radial displacement is given by
\EQ
\label{Delta_disp}
\Delta_{\rm disp}=v_{\rm th}\tau_{\rm th},
\EN
where
\EQ
\label{tau_th1}
\tau_{\rm th}=\frac{D}{u_\theta(\Delta_{\rm disp})}
\EN
is the time
the particle stays within the front side boundary layer and
$u_\theta(\Delta_{\rm disp})$ is the tangential velocity of the flow
in the boundary layer a distance $\Delta_{\rm disp}$ away from the
surface of the cylinder (see \Fig{fig:streamline}).
From \Eqs{utheta}{tau_th1} we can now find the typical 
time the last particle that hits the
front side of the cylinder stays within the 
cylinder boundary layer before it hits the boundary as
\EQ
\label{tau_th2}
\tau_{\rm th}=\frac{BD^{2}}{u_0\Rey^{1/2}\Delta_{\rm disp}}.
\EN
Combining \Eq{tau_th2} with \Eq{v_th}, \Eq{nabla_T} and \Eq{Delta_disp},
and solving for $\Delta_{\rm disp}$, now yields
\EQ
\Delta^2_{\rm disp}=\frac{\Phi B^{2}{\Pr}^{1/3}D^{2}}{6\pi \Rey}\frac{\Delta 
T}{T},
\EN
where we use that $\Rey=u_0D/\nu$.
Having found $\Delta_{\rm disp}$, we proceed by setting 
the two mass fluxes defined in
\Eqs{mdotu}{mdotb} equal to each other and solve for $\Delta x_{\rm max}$ to 
find
\EQ
\Delta x_{\rm max}=\frac{\mean{\rho}}{\rho_0}
\frac{\Rey^{1/2}}{2BD}\Delta^2_{\rm disp}.
\EN
In the above we have used the fact that $\Delta x=\Delta x_{\rm max}$ when 
$\Delta r_f=\Delta_{\rm disp}$.
From \Eq{eta_small}, we now find that for a non-negligible temperature
difference, the capture efficiency for small
Stokes numbers ($\St \lesssim 0.1$) is given by
\EQ
\label{eq:etasmall}
\eta
=\frac{2\Delta x_{\rm max}}{D}
=\frac{\Phi B {\Pr}^{1/3}}{6\pi\Rey^{1/2}}\frac{\Delta T}{T}\frac{\mean{\rho}}{\rho_0}
=\frac{2 K_{tc} B {\Pr}^{1/3}}{\Rey^{1/2}(2+\Lambda)}\frac{\Delta T}{T}\frac{\mean{\rho}}{\rho_0}.
\EN

For laminar flows, \citet{Haugen2010} found that $B$ is independent of
Reynolds number but varies with angular position on the cylinder surface. 
In particular, they found $1/B$ to be zero at the front stagnation
point while the minimum value of B was found to be 0.45 at a position
60 degrees further downstream. 
For the remainder of this paper we
chose $B=1.6$, which is a value that yields
good model predictions. In order to obtain the last part of \Eq{eq:etasmall},
we have used the simplified version of the thermophoretic force term ($\phi$),
but any version can be used here.

It is clear that the above approach, yielding an impaction efficiency due
to thermophoretic forces for small Stokes numbers, is strictly
applicable only when the distance the particle travels
 within the
 boundary layer is less than a fraction $\alpha$ of
 the thickness of the boundary layer itself,
i.e. when
\EQ
\label{applicable}
\frac{\Delta_{\rm disp}}{\delta_{\rm thermal}}
=B^2\sqrt{\frac{2K_{tc}\Pr\Delta T}{(2+\Lambda)T}}
< \alpha.
\EN
We shall later see that the critical value of $\alpha$ is somewhere between
0.5 and 1.

\section{Numerical Methods}
The simulations for the present work are performed using the Pencil Code, an
open source, highly parallelizable code for compressible flows with a
wide range of implemented methods to model different physical effects
\citep{Brandenburg2002,JOSS,pencilpage}. 

The effect of thermophoresis on the impaction efficiency is studied by
releasing a large number of particles upstream of a cylinder in an
established quasi-steady flow field. Every time step, new particles
are inserted at random positions on the particle insertion plane
(shown as a red line in \Fig{fig:sketch}) with a velocity that is equal
to the inlet fluid velocity.  The domain is two-dimensional and has a
width of $6D$ and a length of $12D$, where $D$ is the diameter of the
cylinder.  The flow enters the domain on the left and leaves the
domain through the outlet on the right of the domain. Navier-Stokes
characteristic boundary conditions are applied at both inlet and
outlet to ensure they are non-reflective for acoustic waves
\citep{Poinsot1992}. All other boundaries are periodic. To precisely
represent the cylinder at low computational cost, a cylindrical
overset grid is placed around the cylinder. This cylindrical grid
communicates with the Cartesian background grid via its outer points.
Summation-by-parts is used for derivatives on the surface of the
cylinder, and a Pad\'{e} filter is used to mitigate high frequency
oscillations on the cylindrical grid.  For details concerning the
cylindrical overset grid, the reader is referred
to \citet{Aarnes2018,Aarnes2019,Aarnes2019b}.

The background grid is divided into 288 and 576 cells for the width and
length, respectively, except for the studies at the highest Reynolds
number 1600, where, the resolution is doubled. The cylindrical grid
has 144 cells in the radial and 480 cells in the tangential direction for
the cases with Reynolds numbers up to 400, and double the amount for
the case with a Reynolds number of 1600. Grid stretching in the radial
direction is used to ensure approximately matching cell sizes on the
outer grid points of the cylindrical cells, where the background and
the overset grids communicate.  The code uses a sixth-order finite
difference scheme for spatial discretization and a third-order
Runge-Kutta scheme for temporal discretization. Since the cell size at
the cylinder is much smaller than the general cell size of the
background grid, the time step of the background grid can be a
multiple of the time step of the cylindrical grid, with the
cylindrical flow being updated more often. For details of the particle
tracking scheme, the reader is referred to \citet{Haugen2010} or
\citet{Aarnes2019}.

\section{Simulations}
The inflow has a temperature of 873 K and a density of 0.4 kg/m${^3}$,
corresponding to the density of air at this temperature and a pressure
of 1 bar. The particles have a density of 400 kg/m$^{3}$, leading to a
density ratio $S_{\rho}$ of 1000 based on the fluid density under
inlet conditions.  These values are chosen based on their relevance
for particle deposition on super-heater tubes in thermal power plants,
but the results are nevertheless generic since they are given as
functions of non-dimensional numbers.  The cylinder temperature is
implemented as a Dirichlet boundary condition and set to a fixed
value. The inflow velocity $u_0$ is set so that the flow Reynolds
number $\Rey = u_0D/\nu$ is 100 for the cases not studying the
Reynolds number effects. For the cases studying the Reynolds number
effect, the viscosity is changed to obtain different Reynolds
numbers. To study the same range of Stokes numbers, the particle size
is adjusted accordingly, and the thermal diffusivity is decreased to
achieve a constant Prandtl number.  The different aspects of the
thermophoretic effect are analyzed by changing one critical parameter
at the time, while holding the others constant. These critical
parameters are: 1) Reynolds number, 2) Prandtl number, 3) temperature
difference between fluid and cylinder and 4) conductivity ratio.

The conductivity ratio is given by 
\EQ
\Lambda=\frac{k_p}{k_f}=\frac{k_p}{D_{\rm thermal} c_p 
\rho}=\frac{k_{pf}}{D_{\rm thermal}},
\EN
where $k_{pf}=k_p/(c_p \rho)$. Since we keep the thermal diffusivity
of the fluid constant when changing the conductivity ratio, this means
that the conductivity ratio is essentially changed by changing
$k_{pf}$.

For each of the parameters listed in \Tab{tab:params}, the impaction
efficiencies of particles in the Stokes number range between 0.01 and
10 are obtained from the DNS simulations.
\begin{table}
  \centering
  \caption{Range of parameters studied}
  \label{tab:params}
 \begin{tabular}[]{ll}
  Parameter & Values\\\hline
  Reynolds number [-]& 100, 400, 1600\\
  Conductivity ratio [-]& 1, 12, 144\\
  $\Delta$T [K]& 0, 1, 3, 10, 173, 400\\
 \end{tabular}
\end{table}
For each particle size (Stokes number), a certain number of particles
have to impact the surface to get the statistics required in order to
estimate an accurate impaction efficiency.  Since we already know that
the impaction efficiency decreases with Stokes number, it is therefore
clear that we have to release more small than large particles.  In
particular, for particles with Stokes numbers $> 1$, a total of 15,000 particles
are released for each particle size (Stokes number), while
200,000-400,000 particles are released for particles with $0.1< \St < 1$,
and 2 million particles of each particle size
are released for $\St<0.1$.  The particles
are inserted over several vortex shedding times to mitigate the effect
the instantaneous vortex shedding could have on the results.

All simulations can be described by the $n=10$ unique and independent
variables that are listed in \Tab{tab:indep_params}. From the table we
see that all variables involve a total of $k=4$ different units (m, s, kg and 
K).
\begin{table}
  \centering
  \caption{Independent variables describing the simulations}
  \label{tab:indep_params}
 \begin{tabular}[]{lll}
  Variable & unit & Description \\\hline
  $\alpha_{p}$  & m$^2$/s & Thermal diffusivity of particles \\
$\rho_p$   & kg/m$^3$ & Material density of particles \\
$d_p$      & m        & Diameter of particles \\
$\rho$   & kg/m$^3$ & Material density of fluid \\
$D_{\rm thermal}$ & m$^2$/s & Thermal diffusivity of fluid \\
$u$ & m/s & Velocity of fluid \\
$\nu$ & m$^2$/s & Viscosity of fluid \\
$D$ & m & Diameter of cylinder \\
$T_f$ & K & Far-field temperature of fluid \\
$T_s$ & K & Temperature of cylinder \\
 \end{tabular}
\end{table}
Then, from the Buckingham-Pi theorem, we know that the simulations can
be described by exactly $p=n-k=6$ different dimensionless
numbers. These dimensionless numbers are given in
\Tab{tab:dim_less_num}.
\begin{table}
  \centering
  \caption{Dimensionless numbers describing the simulations}
  \label{tab:dim_less_num}
 \begin{tabular}[]{ll}
  Dimensionless number & Description \\\hline
  $\Lambda=k_{pf}/D_{\rm thermal}$ & Conductivity ratio \\
$\Rey=Du/\nu$ & Reynolds number \\
$\Theta=T_f/T_s$ & Temperature ratio between far-field and cylinder \\
$S_\rho=\rho_p/\rho$ & Density ratio between particle and fluid \\
$\St=S_\rho d_p^2 u/(18\nu D)$ & Stokes number \\
$\Pr=\nu/D_{\rm thermal}$ & Prandtl number \\
 \end{tabular}
\end{table}
In this work, we study the effect of variations in all of these
dimensionless numbers, except for $S_\rho$, which is always kept
constant at a value of one thousand. We know that changing $S_\rho$
means that another dimensionless number, $D/d_p=\sqrt{\Rey S_\rho/(18\St)}$,
will change. The effect of this is that the level of the
interception mode will shift. The interception mode is the mode by
which very small particles intercept the cylinder when only pure
impaction is accounted for. This is important to account for when
comparing simulation results with different values of $S_\rho$, which
for example is done in figure 12b of \cite{Kleinhans2017}.
The simulations presented in the following are listed in \Tab{tab:cases}.

\begin{table}
  \centering
  \caption{Overview of
  	 the simulated cases. The Stokes number is varied 
between 0.01 and 10 for all cases.}
  \label{tab:cases}
 \begin{tabular}{lllll}
\hline
 Sim.&   $\Rey$ &$\Pr$     &$\Delta$T & $\Lambda$\\ 
\hline 
'Base case'    &  100&	0.7&	 -&	  -\\
0     &  100&	0.7&	 173&	  12\\
C1    &  100&	0.7&	 173&	   1\\
C144  &  100&	0.7&	 173&	 144\\
dT3   &  100&	0.7&	   3&	  12\\
dT10  &  100&	0.7&	  10&	  12\\
dT400 &  100&	0.7&	 400&	  12\\
R400  &  400&	0.7&	 173&	  12\\
R1600 & 1600&	0.7&     173&	  12\\
RPv400 &  400&	0.175&	 173&	  12\\
RPv1600& 1600&	0.043&	 173&	  12\\
 \end{tabular}
\end{table}

\section{Results}
In this section we will study the effect of the thermophoretic force
on the impaction efficiency. In particular, we will look at how the impaction
efficiency is affected by changes in
the Reynolds number, temperature difference and conductivity ratio.

\begin{figure}
 \centering
 \includegraphics[width=0.6\textwidth]{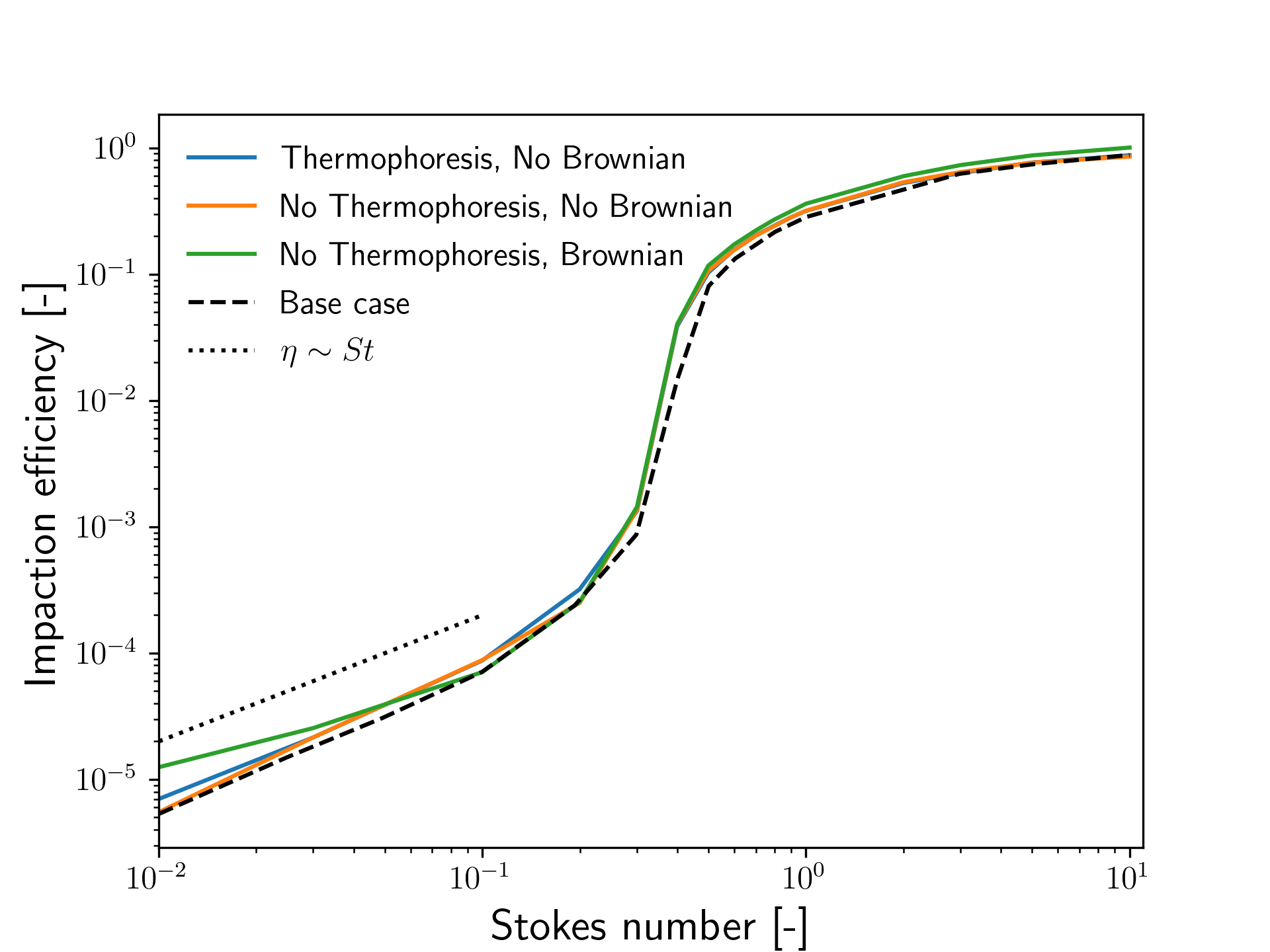}
 \caption{Comparison of $\eta$ for data from \cite{Aarnes2018} with data 
   obtained from an isothermal case with and without thermophoretic force.}
 \label{fig:eff_noeff}
\end{figure}

In their work, \cite{Aarnes2018} used DNS to find the efficiency by
which particles embedded in an isothermal flow impact on a cylinder in
a cross flow. In their study they also used an overset grid, but they
did not consider the thermophoretic force nor did they solve the
energy equation. In the following, we will use their results as a
reference case, from now on called the 'Base case'.
Figure~\ref{fig:eff_noeff} compares the impaction efficiency of the
'Base case' with what is found for the same condition in the current
work. From the figure, we can see that the impaction efficiency shows
a slight decrease with decreasing Stokes number for Stokes numbers
over 1, followed by a steep drop of impaction efficiency in the Stokes
number range between 0.1 and 1. For even smaller Stokes numbers, the
impaction efficiency decreases linearly with decreasing Stokes number.
As expected, our recent simulations of isothermal cases both with
(blue line) and without (orange line) the thermophoretic force
(\Eq{F_th}) included yield the same impaction efficiency profile as
the 'Base case'.  A case with Brownian forces on the particles has
also been performed, and one can see a weak effect of Brownian
forces for the smallest Stokes numbers.  This effect is quite
weak, and, as we shall see later, the thermophoretic force
will have a much stronger effect on the impaction efficiency even for very
small temperature gradients.

\begin{figure}
 \centering
 \includegraphics[width=0.6\textwidth]{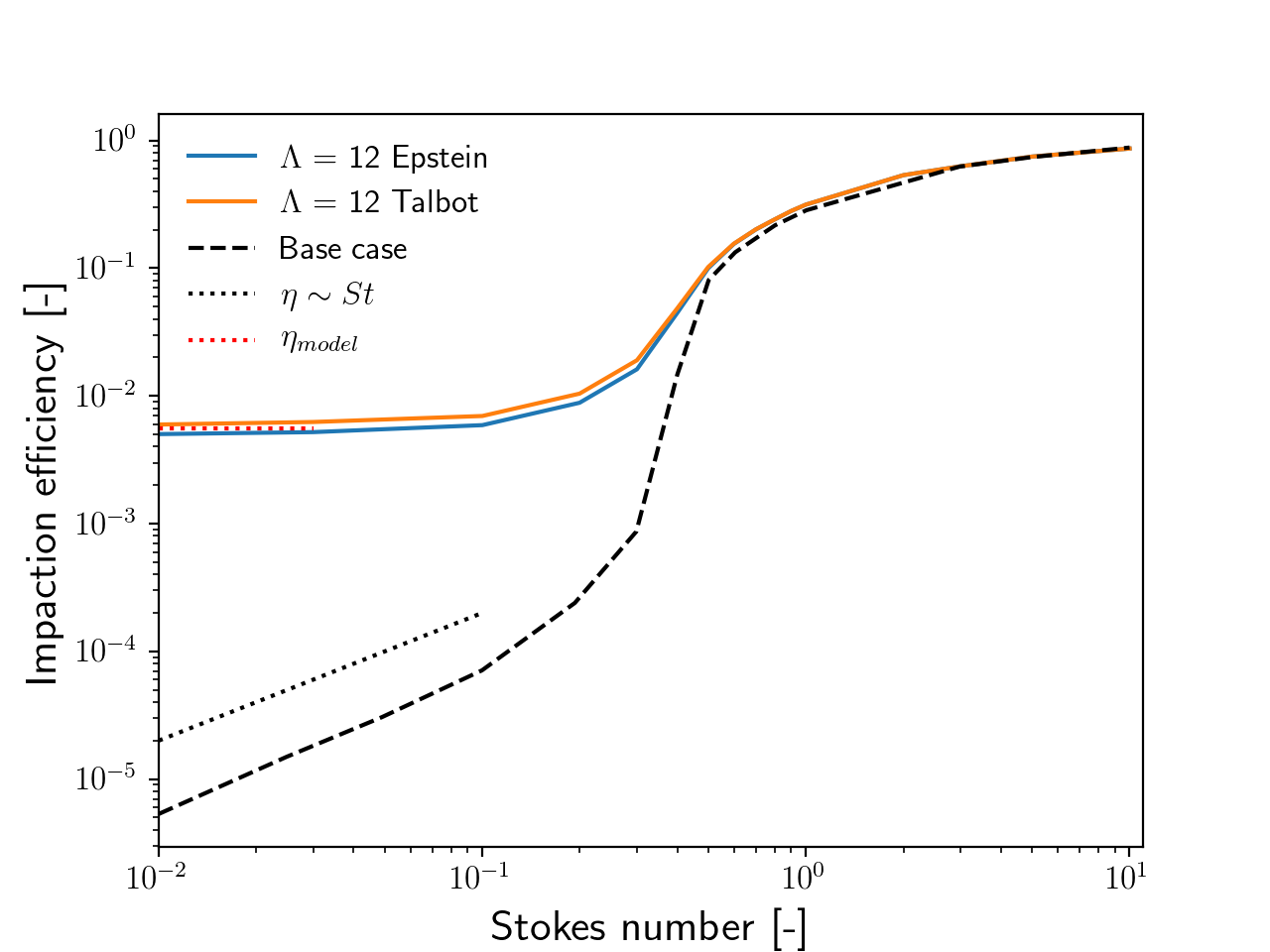}
 \caption{Front side impaction efficiency as a function of Stokes number. For the non-isothermal cases ($\Delta T=173$~K),
   results with thermophoretic factors as given 
   by Epstein and Talbot are compared. The red dotted line corresponds to the 
impaction efficiency predicted by the model for small
 Stokes numbers as given in \Eq{eq:etasmall}.}
 \label{fig:talbot}
\end{figure}

In \Fig{fig:talbot}, the 'Base case' (isothermal) is compared to
non-isothermal cases with thermophoresis.  Simulations with a
temperature difference between the inlet gas and the cylinder surface
of $\Delta T= 173$~K and a particle conductivity ratio of 12 was used
for the thermophoretic cases.  It was shown in \Sec{sec:parteq} that
the models of Epstein and Talbot gave comparable values of
$\Phi$.  The two solid lines in \Fig{fig:talbot} show the
corresponding difference in impaction efficiency.  For the smaller
particles, the impaction efficiency predicted by the model of Talbot
is only about 10\% higher than the one predicted by Epstein, with the
difference disappearing for larger particles.  It is clear from the
figure that the particle impaction is unaffected by the
thermophoretic force for large Stokes numbers, while the
thermophoretic force is dominating the impaction
for small Stokes numbers.  In contrast to what is observed for the
isothermal case, the impaction efficiency becomes independent of the
particle size for small particles.  The theoretical prediction of the
impaction efficiency for small Stokes numbers, as presented in
\Eq{eq:etasmall}, is represented by the red dotted horizontal line in
the figure and one can see that it fits well with the numerical
results for $\St<0.1$.

The effect of different conductivity ratios on the impaction
efficiency is shown in \Fig{fig:cond}.  We see that the
impaction efficiency for small Stokes numbers is higher for lower
values of the conductivity ratio.  This is because a small
conductivity ratio yields a large thermophoretic force term ($\Phi$),
which again results in a large thermophoretic force and hence high
impaction efficiency. This effect can also be seen from
\Eq{eq:etasmall}, where the impaction efficiency for small Stokes
numbers is seen to scale linearly with $\Phi$. When the simplified
expression for $\Phi$ is used, a linear dependence on
$\Phi$ means that the impaction efficiency is inversely proportional
with $(2+\Lambda)$.

For the smallest conductivity ratio (case C1), we find from the
applicability test of our model, as given in \Eq{applicable} and
visualized in \Fig{fig:applicable}, that the thermophoretic force is
so strong that our model is not strictly applicable (i.e. $\Delta_{\rm
  disp}/\delta_{\rm thermal}>\alpha$ when $\alpha\lesssim 0.5$). This
is probably the reason why the modelled impaction efficiency at small
Stokes numbers, as represented by the horizontal dotted lines in
\Fig{fig:cond}, deviates somewhat from the simulated results (solid
lines).  It should be noted that a conductivity ratio of 1 is rather
improbable.  \cite{Zhang1998} give a value for the conductivity ratio
of small char particles of $\Lambda \approx 9$. The conductivity ratio
may, however, be quite different for other solids. In the following we
have somewhat arbitrarily chosen to use $\Lambda=12$ as a baseline
for our simulations. By comparing how well
our model reproduces the simulation results, it seems reasonable to
assume that $\alpha\sim 0.5-1.0$. Based on \Eq{applicable}, we can then
find that for char particles ($\Lambda\sim 10$), the more stringent
value of $\alpha$ (=0.5) results in the model being
applicable as long as $\Delta T$ is less than $\sim 30$\% of the far
field temperature for fluids with $\Pr=0.7$.
For the same conditions and $\alpha = 1.0$,
our model is applicable for all values of $\Delta T$.

\begin{figure}
 \centering
 \includegraphics[width=0.6\textwidth]{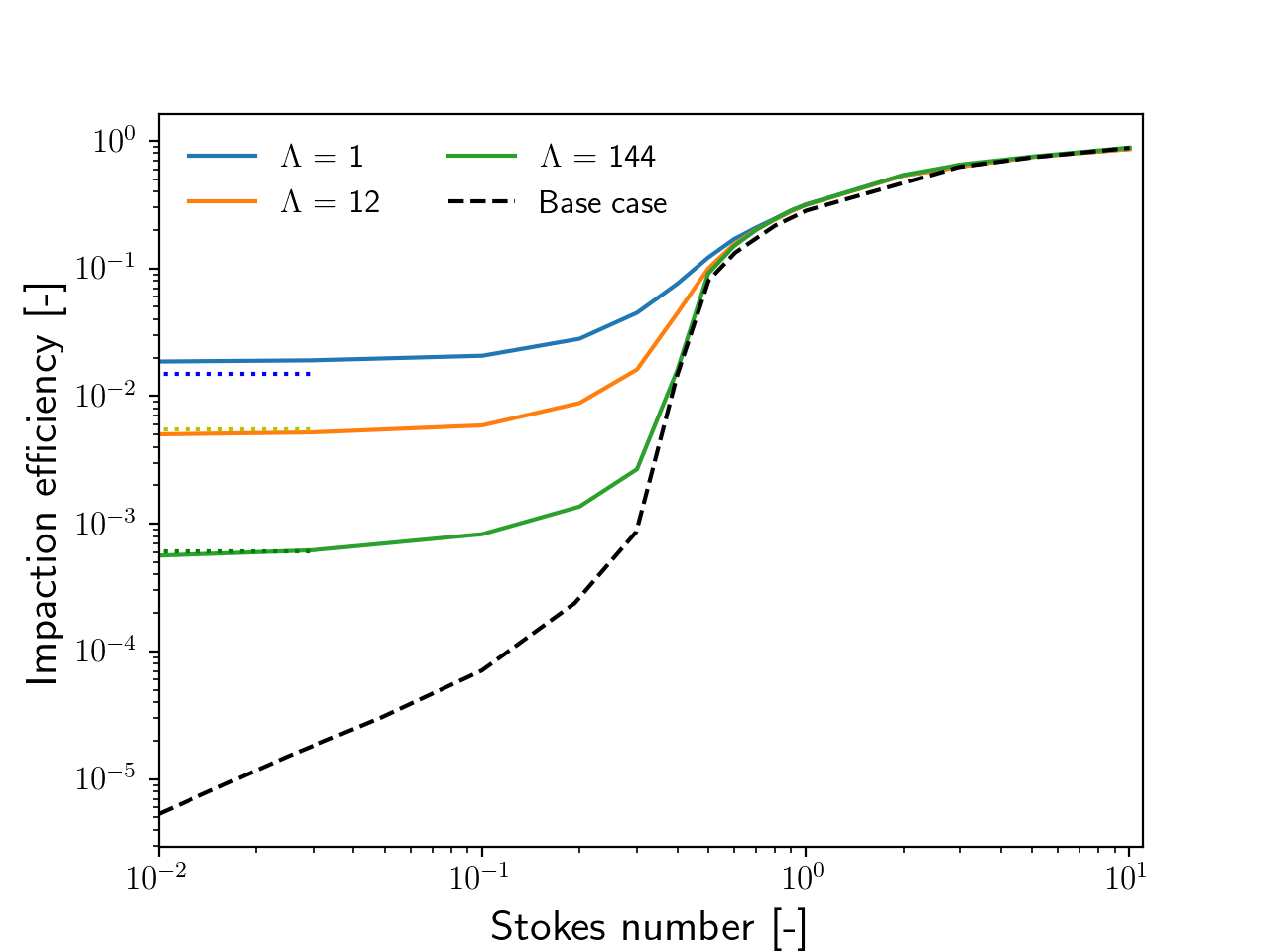}
 \caption{Front side impaction efficiency over Stokes number for
   different conductivity ratios. This corresponds to simulations
   'Base case', 0, C1 and C144 as listed in \Tab{tab:cases}. The
   dotted lines correspond to the impaction efficiency predicted by
   \Eq{eq:etasmall} for small Stokes numbers. }
 \label{fig:cond}
\end{figure}

\begin{figure}
 \centering
 \includegraphics[width=0.6\textwidth]{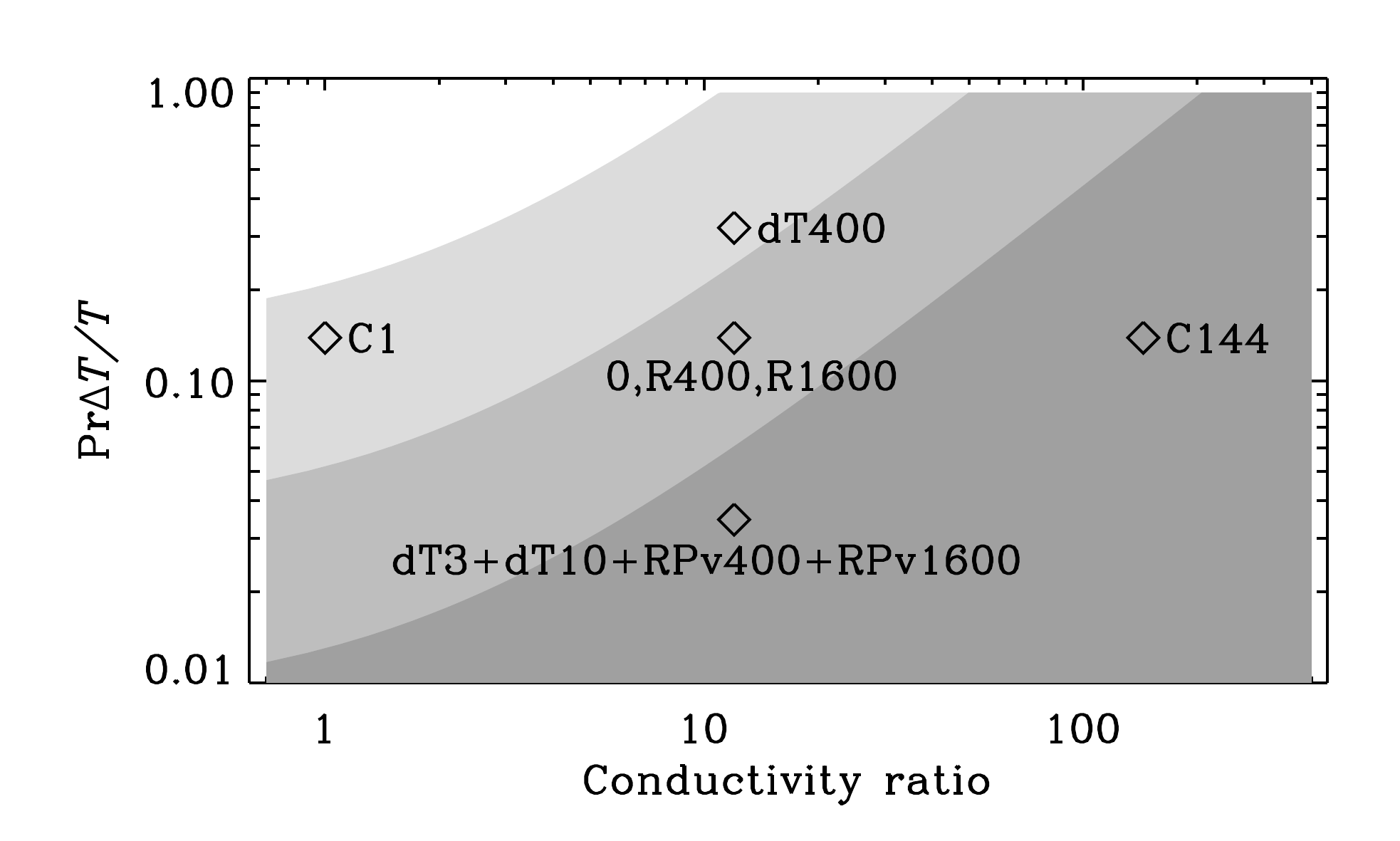}
 \caption{For the simulations marked in the grey area, the model from 
\Eq{eq:etasmall}
is applicable. From lightest to darkest the grey areas correspond to $\alpha=1.0$, 0.5 and 0.25. 
For those that are in the white area, however, \Eq{applicable}
   yields no applicability. The simulations with different Reynolds number and 
constant
   Prandtl number (``R400'' and ``R1600'') are positioned at the same place as
   simulation ``0''.}
 \label{fig:applicable}
\end{figure}

In the left hand panel of \Fig{fig:angle12}, the impaction efficiency
is shown as a function of angular position of impact on the cylinder
for different Stokes numbers for simulations with a conductivity ratio
of $\Lambda=12$. (The center-line is at 270 degrees.) For the largest
Stokes number ($\St=0.9$), all impaction occurs within an angle,
$\theta_{\rm max}$, that is less than 60 degrees from the center
line. This is consistent with the findings of \cite{Haugen2010}
without thermophoresis.  \cite{Haugen2010} showed that $\theta_{\rm
  max}<60$ degrees for $\St<0.9$ for pure inertial impaction (see also
\Fig{fig:angle12_noth}). However, we see that when thermophoresis is
accounted for, the particles with smaller Stokes numbers impact the
entire frontal surface of the cylinder. By increasing the strength of
the thermophoretic force, which is here done by decreasing the
conductivity ratio to unity, we see from the right hand panel of
\Fig{fig:angle12} that the angular position of impaction becomes even
more uniform for the smaller particles.

\begin{figure}
  \centering
  \includegraphics[width=0.49\linewidth]{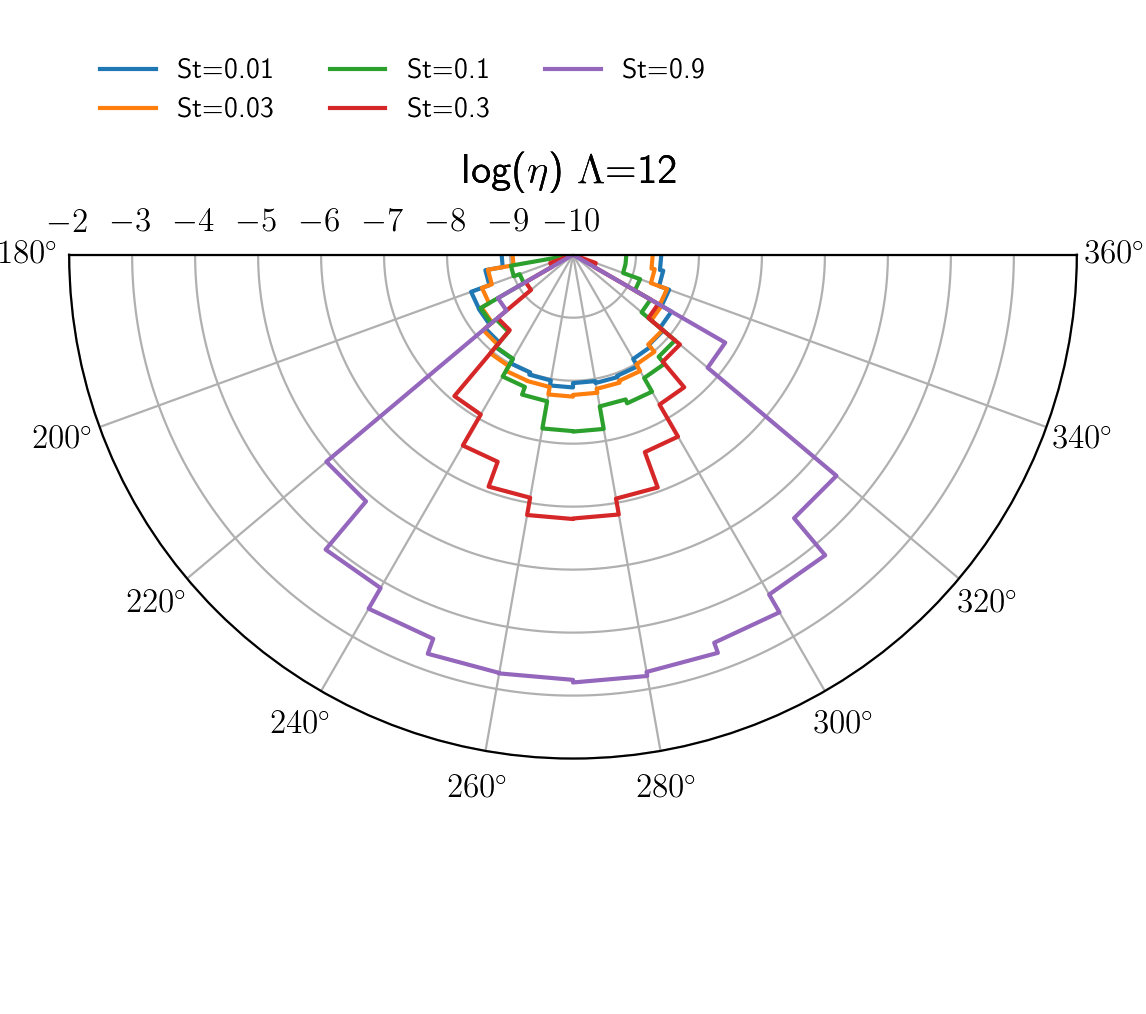}
  \includegraphics[width=0.49\linewidth]{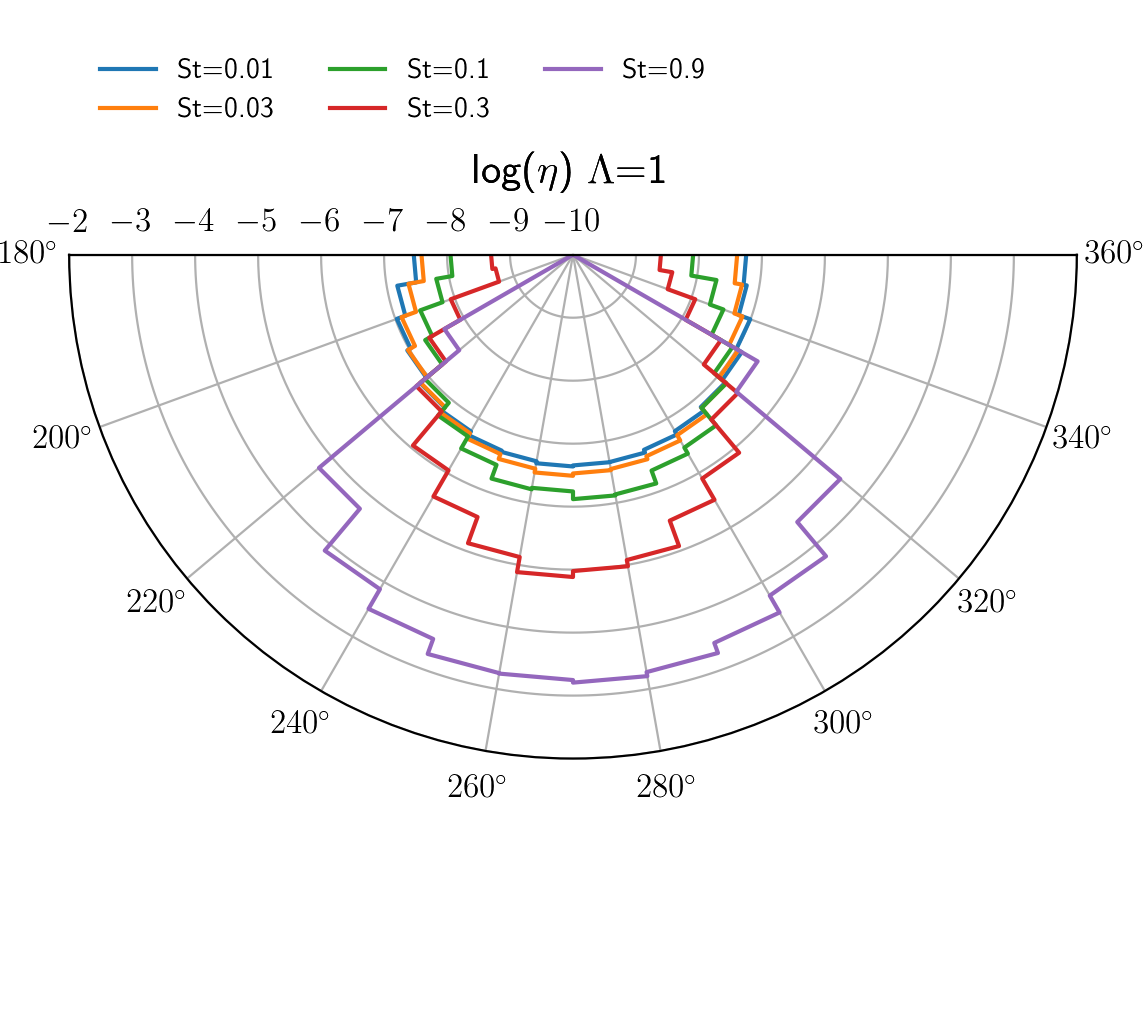}
  \caption{Front side impaction angle for $\Lambda$=12 (left panel) and 
$\Lambda$=1 (right panel). Simulations 0 and C1, respectively.}
\label{fig:angle12}
\end{figure}

\begin{figure}
  \centering
\includegraphics[width=0.49\linewidth]{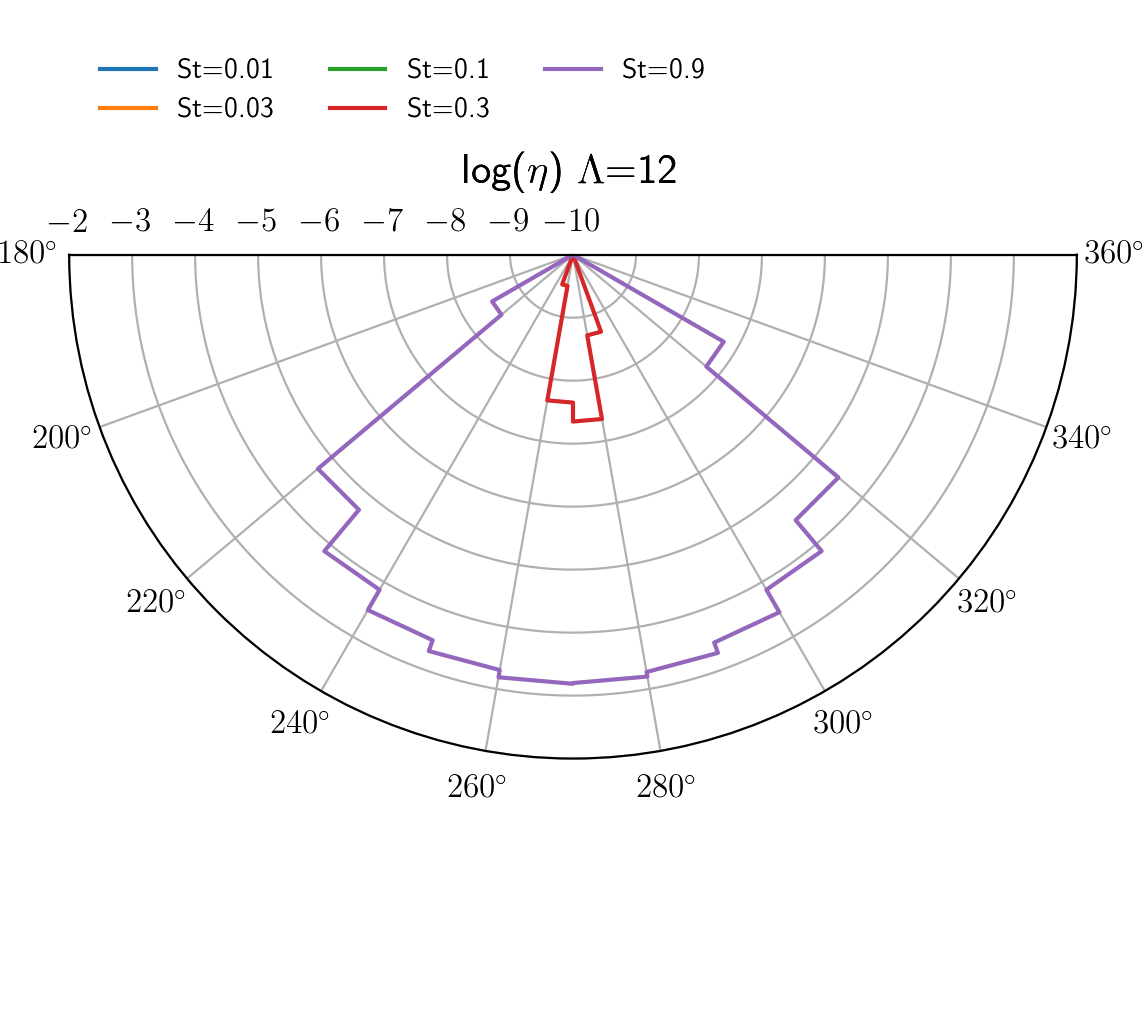}
  \caption{Front side impaction angle for $\Lambda$=12 with no thermophoresis.}
\label{fig:angle12_noth}
\end{figure}

\begin{figure}
 \centering
 \includegraphics[width=0.6\textwidth]{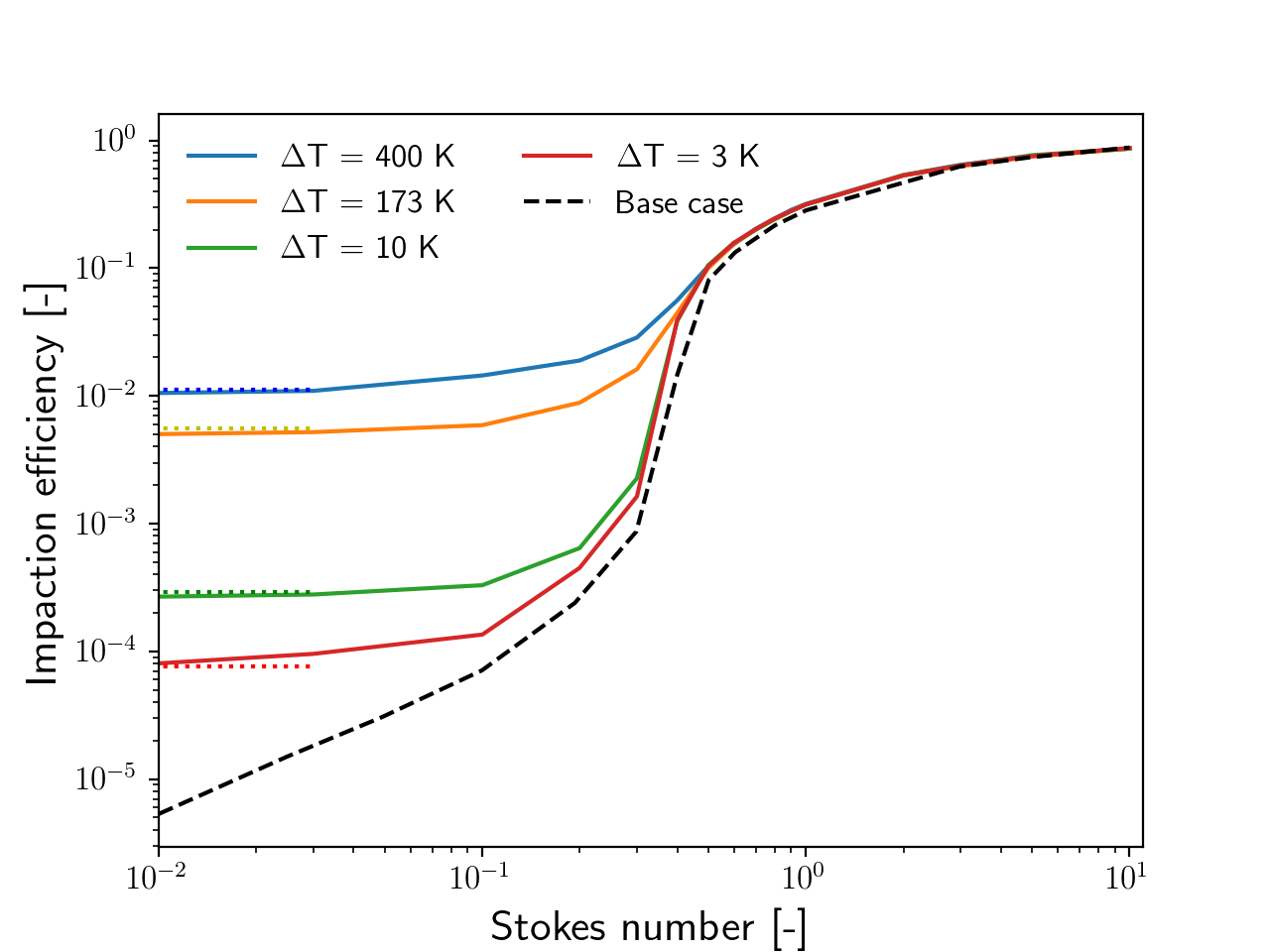}
 \caption{Impaction efficiency over Stokes number for different cylinder 
   temperatures (simulations 0, dT3, dT10 and dT400 as listed in 
\Tab{tab:cases}).
   A high temperature difference increases $\eta$ for 
small particles. The dotted lines correspond to the impaction 
efficiency predicted by \Eq{eq:etasmall} for small
 Stokes numbers.}
 \label{fig:T}
\end{figure}

Since the gas is an ideal gas, the cylinder is surrounded by a
boundary layer of densified gas with a significant temperature
gradient for the cases with large temperature differences. A larger
temperature difference yields a stronger thermophoretic force, which
again results in a higher impaction efficiency for small particles. As
can be seen from \Fig{fig:T}, the model results (dotted lines) fit the
simulation results (solid lines) well for all temperature differences
studied here ($3\; \mbox{K}<\Delta T<400$ K).  From
\Fig{fig:applicable} we do see, though, that $\Delta T=400$~K is close
to the limit of the applicability of the model.

We will now continue by studying the effect of Reynolds number on the
impaction efficiency. In this work, we increase the Reynolds number by
decreasing the viscosity. While doing this, we must also increase the
resolution in order to properly resolve the boundary layer, which is
thinner for higher Reynolds numbers. If the thermal diffusivity is
changed linearly with viscosity, the Prandtl number is kept constant.
From \Figs{fig:re}{fig:re_pr_var} we see that increasing the flow
Reynolds number results in higher front side impaction efficiencies
for intermediate Stokes numbers in the range $0.1 < \St < 1$.
Qualitatively, this is consistent with the findings of
\cite{Haugen2010} obtained for isothermal cases.  From \Fig{fig:re},
which is obtained by maintaining a constant Prandtl number, we see
that there is a clear but not dramatic Reynolds number effect for
small Stokes numbers. This is supported by the model in
\Eq{eq:etasmall}, which shows that the impaction efficiency for small
Stokes numbers should be inversely proportional to the square root of
the Reynolds number. This is represented by the horizontal dotted
lines in the figure, which accurately reproduce the DNS results.
The slight increase in impaction efficiency when the Stokes number is
decreased from 0.03 to 0.01 for Re=1600 is due to poor statistics.
It can also be mentioned that, since higher Reynolds numbers are obtained
by decreasing the viscosity, the particles for low Stokes numbers
become quite small. Unlike for isothermal cases, particle size does
not, however, play any role in the impaction efficiency for small
Stokes number when their impaction is controlled by the thermophoretic
force.

If the thermal conductivity is not changed when changing viscosity,
the Prandtl number is decreased for increasing Reynolds numbers, which
is the case for the simulations shown in \Fig{fig:re_pr_var}. Here we
see that the difference in impaction efficiency for the smaller Stokes
numbers is larger than for the case with constant Prandtl number. By
comparing the solid and the dotted lines, we see that the DNS
results (solid lines) follows the model predictions (dotted lines) nicely.

\begin{figure}
 \centering
 \includegraphics[width=0.6\textwidth]{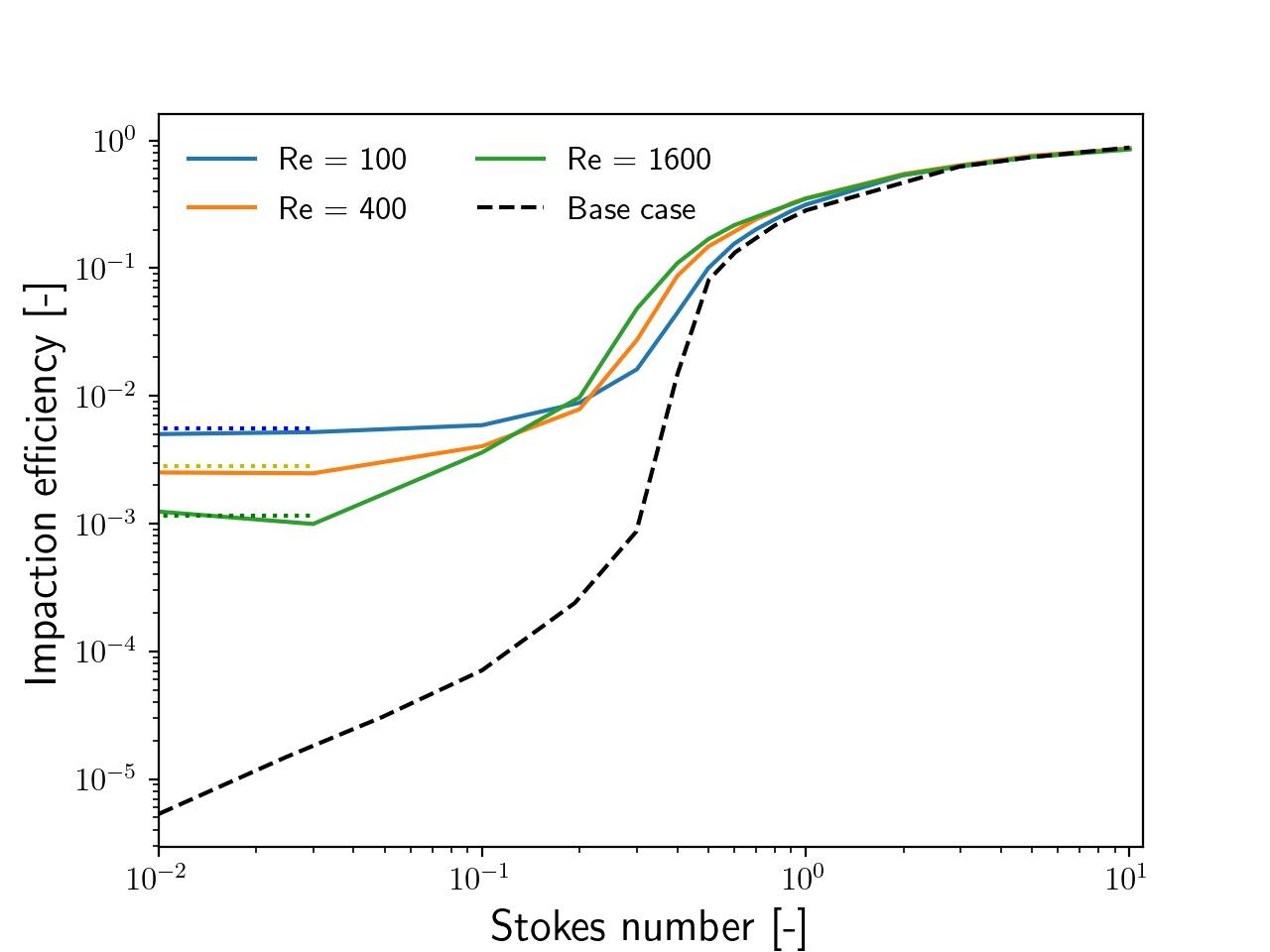}
 \caption{Front side impaction efficiency over Stokes number for different flow 
Reynolds 
numbers. Simulations ``0'', ``R400'' and ``R1600''. Higher Reynolds numbers 
result in higher $\eta$ for 
medium Stokes numbers, while for low Stokes 
numbers $\eta$ is decreased. The dotted lines correspond to the impaction efficiency predicted by 
\Eq{eq:etasmall} for small
 Stokes numbers.}
 \label{fig:re}
\end{figure}
\begin{figure}
 \centering
 \includegraphics[width=0.6\textwidth]{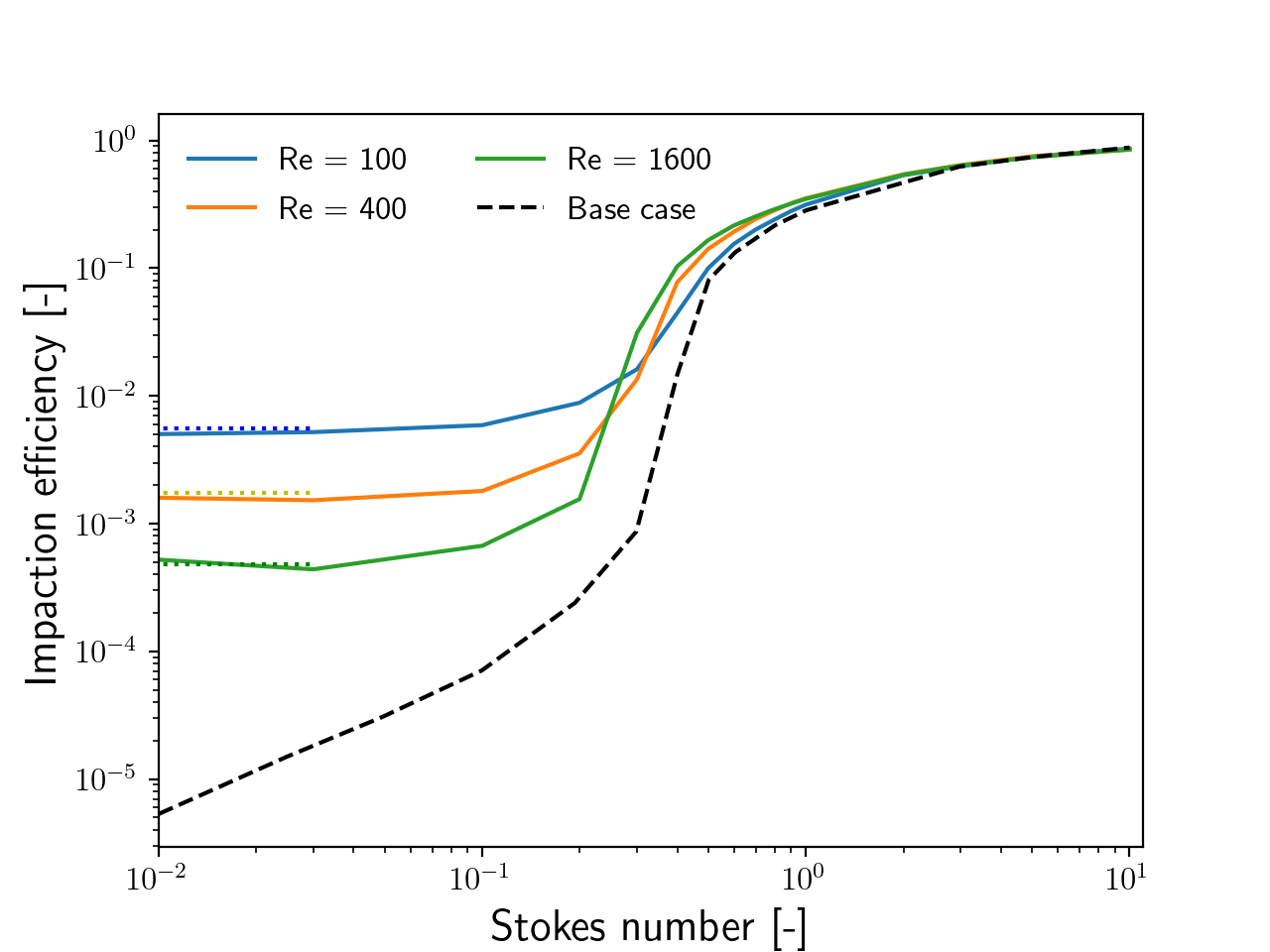}
 \caption{Front side impaction efficiency over Stokes number for different flow 
Reynolds 
numbers. The Prandtl number is inversely proportional to the Reynolds 
number, such that it equals 0.7, 0.175 and 0.043 for $\Rey=$100, 400 and 1600,
respectively. Simulations ``0'', ``RPv400'' and ``RPv1600''.
Higher Reynolds numbers result in higher $\eta$ for 
medium Stokes numbers, while for low Stokes 
numbers $\eta$ is decreased. The dotted lines correspond to the impaction efficiency predicted by 
\Eq{eq:etasmall} for small
 Stokes numbers.}
 \label{fig:re_pr_var}
\end{figure}

\section{Conclusions}
The effect of thermophoresis on the impaction efficiency of particles
on a cylinder is studied for different values of the Reynolds number,
conductivity ratio, temperature difference and particle Stokes
number. Compared to the case where thermophoresis is neglected, the
impaction efficiencies become independent of Stokes number for small
Stokes numbers ($\lesssim 0.1$), corresponding to small particles. For
small Stokes numbers, the impaction efficiency is larger for low
conductivity ratios, high temperature differences and low Reynolds
numbers.  Thermophoresis has an insignificant effect on particles with
Stokes numbers larger than $\sim 0.5$, but is often dominating the
impaction rate for $\lesssim 0.3$.

An algebraic model (see \Eq{eq:etasmall}) that predicts the impaction
efficiency due to thermophoresis has been developed based on
fundamental principles.  The developed model is valid as long as the
thermophoretic force is not too strong (see \Eq{applicable} and
\Fig{fig:applicable}). Outside its range of validity, a reliable model
does not yet exist. This should be the focus of future research.

\section*{Acknowledgements} 
This research was supported by The GrateCFD project [grant
  267957/E20], which is funded by: LOGE AB, Statkraft Varme AS, EGE
Oslo, Vattenfall AB, Hitachi Zosen Inova AG and Returkraft AS together
with the Research Council of Norway through the ENERGIX
program. Computational resources was provided by UNINETT Sigma2 AS
[project numbers NN9405K].  We would also like to acknowledge that
this research was partly funded by the Research Council of Norway
(Norges Forskingsr\aa d) under the FRINATEK Grant [grant number
  231444] and by the grant
``Bottlenecks for particle growth in turbulent aerosols''
from the Knut and Alice Wallenberg Foundation, Dnr.\ KAW 2014.0048.

\bibliography{Main}
\end{document}